\input mnrass.sty
\pageoffset{-2.5pc}{0pc}

 

\Autonumber  


\pagerange{000--000}
\pubyear{1996}
\volume{000}

\begintopmatter  

\title{A critical investigation on the discrepancy between the observational
and theoretical Red Giant luminosity function \lq{Bump}\rq }

\author{ Santi Cassisi$^{1,2}$ \& Maurizio Salaris$^{3,4}$}

\affiliation{$^1$Universit\'a degli studi de L'Aquila, Dipartimento di Fisica,
Via Vetoio, I-67100, L'Aquila, Italy}
\affiliation{$^2$Osservatorio Astronomico di Collurania, Via M. Maggini,
 I-64100, Teramo, Italy - E-Mail: cassisi@astrte.te.astro.it}
\affiliation{$^3$Max-Planck-Institut f\"ur Astrophysik, D-85740,
Garching, Germany - E-Mail: maurizio@MPA-Garching.mpg.de} 
\affiliation{$^4$Centre d'Estudis Avan\c cats de Blanes, C.S.I.C.,
E-17300, Blanes, Spain}

\shortauthor{S.Cassisi \& M.Salaris}
\shorttitle{Red Giant luminosity function \lq{Bump}\rq}



\abstract
\tx
New theoretical evaluations of the RGB luminosity function 'bump' and 
the ZAHB luminosity
covering the range of metallicities typical of galactic globular cluster
are presented. The variation of the theoretical RGB bump and ZAHB levels due to 
the metallicity, original helium content, mixing length value, age,
mass loss, bolometric corrections, opacities and equation of state  
adopted in the evolutionary models is also discussed.
These new prescriptions have been taken into account for
casting light on a longstanding astrophysical problem
connected with the Red Giant Branch evolutionary phase, namely
the discrepancy between the observational and the theoretical
luminosity of RGB bump.
A sample of globular
clusters with accurate evaluations of the bump luminosity
and spectroscopical metallicity determinations
has been selected. The Zero Age
Horizontal Branch luminosity at the RR-Lyrae instability strip
has been evaluated as accurately as possible, and
the observational luminosity difference between the RGB bump
and the ZAHB has been compared with the theoretical values.
It is shown that there is no significant disagreement between observations
and canonical stellar models. The possible applications of this result
are also briefly discussed.


\maketitle  

\section{Introduction}

\tx 

The existence of a well developed and populated Red  Giant  Branch
(RGB) is  one of  the  main  characteristics  of  the  Color
Magnitude Diagrams (CMDs) of  galactic Globular Clusters
(GCs). While
the $\rm T_{eff}$ location of the RGB  provides  a  means  to  calibrate  the
mixing length  parameter  for GC stars
(Chieffi,  Straniero  \&
Salaris 1995, Salaris \& Cassisi 1996 hereinafter Paper I),  the
Luminosity  Function
(LF) of the RGB of GCs
is a useful tool to test  the  inner chemical  composition
stratification,  since  the   hydrogen   abundance   outside   the
degenerate helium core sampled  by  the  thin hydrogen burning
shell, affects the rate of evolution during the RGB phase.
In particular, due to the fact that the shell is extremely  narrow
in mass (thickness of the order of 0.001-0.0001 $M_{\odot}$), its
crossing any discontinuity in the hydrogen profile leads  to  a
temporary drop in the luminosity and therefore to a characteristic 'bump' in the
observed differential LF.

It has been  known for a long time (see  e.g.  Thomas  1967,
Iben 1968, Renzini  \&  Fusi  Pecci  1988,
Castellani, Chieffi  \&  Norci  1989, Bono \& Castellani 1992)  
that  theoretical  RGB  LFs
display a characteristic bump due to  the passage of the
thin  hydrogen  burning
shell through the composition discontinuity left by  the
deepest penetration of  the  convective  envelope. In spite of this,
the first clear identification of
the RGB bump in the observed LF of GCs is relatively recent; it 
goes back to the  work  by
King et al. (1985) on 47 Tuc, and only more recently Fusi Pecci  et  al.
(1990, hereinafter FP90 -
11 clusters), Bergbush  (1993  -  NGC288),  Sarajedini  \&
Norris (1994 - 47  Tuc  plus  other  5  clusters),  Sarajedini  \&
Forrester (1995 - NGC6584), Brocato et al. (1996 - NGC5286, NGC6266,
NGC6934, NGC6981) have extended the sample of GCs with a
detected RGB bump.

The reason for the difficulty in detecting the bump from
GCs photometry  is the need of very large  RGB  star  samples
in order to obtain
a firm identification from the observed LF.
According to the discussion in FP90 at least 1000-1500 stars in the upper 4
magnitudes of the LF are required for a clear identification of the  bump
in metal poor GCs; this star sample is larger than the  ones  currently
available for the best studied metal poor clusters. At higher metallicities
the extension in luminosity of the bump is larger, and it is
also shifted to lower luminosity, in more populated RGB regions, 
so it is easier to detect.

In order to avoid uncertainties due  to  the  calibration  of  the
observational data and to  the  cluster  distance  modulus  it  is
more reliable to consider, when comparing theoretical RGB bump
luminosities    with    the    observations,     the     parameter
$\Delta\rm {V}^{bump}_{hb}=(V_{bump}-V_{hb})$ (FP90)  defined   as   the
difference in visual magnitude between the RGB bump and  the  Zero
Age Horizontal Branch (ZAHB) level.

FP90 tried  to use the bump luminosity as a standard candle, in order
to calibrate the relation between luminosity of the HB and
metallicity. Since the bump luminosity depends also on the age of the
stellar populations, and both bump and ZAHB are affected by the He
abundance,  
they explored different scenarios about the age
of the GC population and the He content. They
compared  the
observed values  of  $\Delta\rm {V}^{bump}_{hb}$  from their sample with  the
theoretical  ones
obtained from RGB models by Rood \& Crocker (1989) and unpublished HB
models by Rood (transformed to the observational plane by adopting the
Bell \& Gustafsson 1978 and Kurucz 1979 transformations);
they derived that the best agreement was obtained by 
assuming a constant age (t=15 Gyr) and a constant He abundance (Y=0.23),
 or a constant age (t=15 Gyr) and an He abundance scaling as $dY/dZ=3$
(according to Peimbert \& Torres-Peimbert 1977).
In these two cases the results of the comparison were:

\noindent
i) the run  of  the  theoretical  relation  with  respect  to  the
metallicity is in very good agreement with the observations.

\noindent
ii) the absolute value of the theoretical relation  is  0.4  mag  too
bright.

\noindent
A possible interpretation of this discrepancy has been presented
by Alongi et al. (1991); they interpreted the detected difference in the
zero-point as a limitation of the
standard stellar models, and proposed undershooting from the base of
the formal boundary of the convective envelope to reconcile theory
and observations.
More recently Straniero, Chieffi \&  Salaris  (1992)  and  Ferraro
(1992)  pointed  out  that the proper
inclusion   of the  $\alpha$-element enhancement in  the  original  chemical
composition of the GC stars (see e.g. Wheeler et al. 1989)
could help in reducing only  partially the discrepancy of the
zero-point without invoking undershooting, but until now the question
remains unsettled.

In this paper we present new  theoretical determinations of  the
RGB bump and ZAHB luminosities covering the range of metallicities
typical of galactic GCs, and
discuss the influence of different physical and chemical inputs
adopted in the model computations on the value of $\Delta\rm {V}^{bump}_{hb}$.
Moreover, we reexamine   critically   the observational    values    of
$\Delta\rm {V}^{bump}_{hb}$, and by comparing
theory with available observational data, we will show
that actually  a  significant discrepancy  between
standard stellar models and observations does not exist.

The plan of the paper is as follows:
the theoretical models  are  presented  in  section  2,  while
section 3 deals with the  discussion of  the  observational
data  and  their  comparison  with   the   theoretical   values   of
$\Delta\rm {V}^{bump}_{hb}$; conclusions follow in section 4.
 
\section{\bf Theoretical determination of the HB and RGB bump luminosity
 level}

\tx
In order to compare the observational values of $\Delta\rm {V}^{bump}_{hb}$
with the theoretical prescriptions, we have computed canonical evolutionary 
models (no diffusion, no undershooting from the base of 
the convective envelope) 
of stars with masses of $0.75M_\odot$, $0.80M_\odot$ and
$0.90M_\odot$ - covering a range of masses presently evolving
along the RGB of GCs corresponding to ages between approximately 22 Gyr
and 12 Gyr - metallicities Z=0.0001, 0.0003, 0.0006, 0.001, 0.003,
0.006, and Y=0.23.  We have interpolated between these models for each
metallicity, in order to obtain the value of $\rm V_{bump}$ corresponding
to an average age t=15 Gyr for the clusters (the choice of
this average age comes out from the extensive analysis about GCs ages
by Chaboyer, Sarajedini \& Demarque 1992, and Salaris, Chieffi \&
Straniero 1993). Since until now it is not clear if an age spread 
really exists in the bulk of the GCs population, we prefer to
adopt this conservative hypothesis about the age of the clusters,   
at least for the purposes of this paper (but see also section 4).
This corresponds to our reference scenario, in which all the GCs are
coeval and with the same He content. 

The He core mass and the surface He abundance at the He ignition 
corresponding 
to the same age of 15 Gyr have
been adopted in order to compute Zero Age Horizontal Branch (ZAHB)
models of different total masses for each value of Z. 
As for the calibration of the superadiabatic envelope convection,
the mixing length calibration described in Paper I,
obtained by fitting the empirical $\rm T_{eff}$ values of
GC RGBs obtained by Frogel et al. (1983), has been adopted.
The evolutionary tracks and the ZAHB models have been translated to
the observational plane by adopting the Kurucz (1992) transformations.
The theoretical values of $\Delta\rm {V}^{bump}_{hb}$ have then been 
computed by considering the difference in $V$ magnitudes between the 
mean $\rm V$ magnitude of the RGB bump region, and $\rm V_{zahb}$
taken at $\rm log(T_{eff})=3.85$, corresponding approximatively to the
average temperature in the RR Lyrae instability strip.

The theoretical relation
$\Delta\rm {V}^{bump}_{hb}$ - [M/H], obtained by means of the models previously
described, is our
$reference$ relation that will be compared with the observations.
In Table 1, the value of $\rm V_{bump}$, $\rm V_{zahb}$ at $\rm log(T_{eff})=3.85$
and $\Delta\rm {V}^{bump}_{hb}$ corresponding to our reference case, are
listed for the different metallicities considered.

\table{1}{S}{\bf Table 1. \rm Luminosity level of
the Horizontal branch at $\rm log(T_{eff})=3.85$, of the RGB bump
and $\Delta\rm {V}^{bump}_{hb}$ obtained by adopting our reference
models (see text),
Y=0.23 and t=15 Gyr.} 
{\halign{%
\rm#\hfil&\hskip10pt\hfil\rm#\hfil&\hskip10pt\hfil\rm\hfil#&\hskip10pt\hfil\rm\hfil#\cr
[M/H] & $\rm M^{zahb}_V$  & $\rm M^{bump}_V$ & $\Delta\rm {V}^{bump}_{hb}$ \cr 
\noalign{\vskip 10pt}
 -2.348 &  0.557 &  -0.332 & -0.889  \cr 
 -1.871 &  0.636 &  -0.051 & -0.687  \cr
 -1.569 &  0.679 &   0.182 & -0.497  \cr   
 -1.347 &  0.715 &   0.355 & -0.360  \cr
 -0.869 &  0.823 &   0.854 &  0.031  \cr
 -0.567 &  0.941 &   1.332 &  0.391  \cr}}
All the theoretical models have
been computed adopting the FRANEC evolutionary code (see Chieffi \&
Straniero 1989). The OPAL opacity tables (Rogers \& Iglesias
1992, Iglesias, Rogers \& Wilson 1992) for $T>10000K$ and the
Alexander \& Ferguson (1994) opacities for $T<10000K$ have been used. 
Both high and low temperature opacity tables are computed adopting the
solar heavy elements distribution (Grevesse 1991).
The electronic
conduction is treated according to Itoh et al. (1983). 
The equation of state (EOS) by Straniero (1988) has
been used, supplemented by a Saha EOS at lower temperatures, as
described by Chieffi \& Straniero (1989).
\figure{1}{S}{100mm}{\bf Figure 1. \rm {\it Top}:
Absolute luminosity of field and globular cluster RR-Lyrae
stars versus  metallicity provided by Clementini et al. (1995).
The luminosity level of the ZAHB
at the average temperature of the instability strip
for two different assumptions concerning the equation of state
are also displayed. The solid line was evaluated taking into account
the EOS provided by Straniero (1988), whereas the dotted line 
is based on the EOS recently provided by the OPAL group;
{\it Bottom}: the same as the top panel but the absolute
luminosities of the RR Lyrae variables have been estimated by
adopting a different value for the conversion factor p
(i.e. the parameter used for trasforming the radial velocities into
pulsational velocities).}

Before star\-ting the compa\-ri\-son between our theo\-reti\-cal
determinations for $\Delta\rm {V}^{bump}_{hb}$ and the observations, 
we have also analyzed in detail (computing additional stellar models) 
the dependence of this quantity 
on other different physical and chemical inputs adopted in computing
stellar models and on the transformations used to transfer
the models from the theoretical plane to the observational one.
Whereas in literature it is easy to find a lot of exhaustive investigation
on the influence on the luminosity level of the HB due to the various
inputs adopted in evolutionary computations (Caloi, Castellani \& Tornamb\'e
1978, Dorman, Rood \& O'Connell 1993 and references therein),
a detailed investigation concerning the properties of the RGB bump 
and therefore of the $\Delta\rm {V}^{bump}_{hb}$ is, till now, still
lacking.
In the following we will discuss the variation of $\Delta\rm {V}^{bump}_{hb}$
induced by changes of the following parameters:
\medskip
\noindent
1) the metallicity and the distribution of the heavy elements;
\smallskip\noindent
2) the abundance of Helium Y;
\smallskip\noindent
3) the mixing length parameter;
\smallskip\noindent
4) the age of the stellar population;
\smallskip\noindent
5) mass loss during the RGB evolution;
\smallskip\noindent
6) the equation of state;
\smallskip\noindent
7) the opacity;
\smallskip\noindent
8) bolometric corrections.
\medskip
\noindent

\subsection{Metallicity and heavy elements distribution.}
\tx

The luminosity of the RGB bump is strongly affected by a variation of the
global amount of heavy elements.
As a matter of fact a metallicity increase causes a decrease of the luminosity
level of the bump.
 This occurrence is due to the
larger extension in mass of the convective envelope during the first dredge up,
related to the larger opacity and then, larger values of the radiative
gradient with respect to the adiabatic one. Therefore the H
discontinuity, 
located deeper and deeper with
increasing  metallicity, is reached by the H shell burning earlier during
the RGB evolution.
Fitting the values of $\rm V_{bump}$ in Tab. 1 as a function of 
$\rm [M/H]=log(M/H)_{star}-log(M/H)_\odot\approx{log(Z)}+1.65$ (where M here is the
global metal abundance and Z the global heavy elements fraction) 
for t=15 Gyr 
yields:
 
$$\rm M^{bump}_V= 2.212 + 1.768\cdot[M/H] + 0.294\cdot[M/H]^2\,\,\,\,\,(1)$$

\noindent
with a r.m.s.=0.02 mag.

Also the luminosity of the ZAHB is strongly affected by the heavy
elements abundance, due to the variation of the He core
mass at the flash with the metallicity. The higher the metallicity,
the lower the He core mass and the ZAHB luminosity.
From our reference models we derived the following relation between
$\rm M^{zahb}_V$ at $\rm log(T_{eff})=3.85$ and [M/H]: 

$$\rm M^{zahb}_V= 1.129 + 0.388\cdot[M/H] + 0.063\cdot[M/H]^2\,\,\,\,\,(2)$$ 

\noindent
with a r.m.s.=0.011 mag.

In Fig.1 a to b, we compare this relation with the most recent
determination of the relation between absolute luminosity of field and globular
cluster RR-Lyrae stars and metallicity, published by Clementini
et al. (1995), based on
new spectroscopical determinations of metallicity and on Baade
Wesselink estimates of the absolute magnitudes of the variables. 
It is worth noting that in Fig.1, only observational data 
for supposedly unevolved RR Lyrae stars have been considered and therefore
they should provide a good estimation of the ZAHB luminosity at the
RR Lyrae instability strip. Moreover,
Clementini et al. (1995) provide two different tabulations
(their Tab.21, columns 5 to 6) 
concerning the absolute luminosity of their RR Lyrae stars sample,
corresponding to two different prescriptions for the conversion
factor p between observed and true pulsational velocity. In panels a to b we
have displayed their data corresponding respectively to columns 6 and
5 of their Tab. 21.  
We have derived the global metallicity [M/H] for the stars in their
sample by assuming the [Fe/H] and [$\alpha$/Fe] values given in the
paper, and also the errors on the observational determinations of 
metallicity and luminosity come
from their paper. 
As is evident from the figure, our theoretical relation is in agreement
- for both choices concerning the conversion factor p - 
with the observational data.

From relations (1) and (2) we derive:

$$\Delta\rm {V}^{bump}_{hb}= 1.083 + 1.380\cdot[M/H] + 0.231\cdot[M/H]^2\,\,\,\,\,(3)$$

\noindent
with a r.m.s.=0.023 mag.

These relations, and in particular the third one, that we will use for
the comparison with the observations, have been derived from stellar
models computed adopting a scaled solar heavy elements
distribution. As well known the original chemical composition of
GCs stars is characterized by $[\alpha/Fe]>0$ (see
e.g. Wheeler et al. 1989); Salaris, Chieffi \& Straniero (1993) have
demonstrated that the evolution of low mass low metallicity stars with
an $\alpha$-enhanced heavy elements distribution and a fixed global metallicity
[M/H], is very well reproduced by scaled solar models with the
same value of [M/H].   

We have verified that the same holds if we adopt the recent $\alpha$-enhanced
OPAL and Alexander \& Ferguson molecular opacities (see Salaris
et al. 1996). In particular the value of $\Delta\rm {V}^{bump}_{hb}$,
obtained using these opacity tables for
a fixed [M/H],  is coincident - within 0.01 mag - with the one derived
from the scaled solar models with the same [M/H].

\subsection{The Helium abundance.}
\tx

The influence of a variation of the original He
content on the value of $\Delta\rm {V}^{bump}_{hb}$ has been also tested.
As well known, increasing
the original He content increases the ZAHB luminosity, due to the more
efficient energy generation in the H-burning shell. At the same time,
also the level of the bump is shifted to higher luminosity, and the
net effect is a slight reduction of the $\Delta\rm {V}^{bump}_{hb}$.
For each $\Delta{Y}=+0.01$ we derive from our models a reduction
of 0.011mag of the
$\Delta\rm {V}^{bump}_{hb}$.

\subsection{The mixing length parameter.}
\tx

The mixing length parameter {\sl ml} is one of the free parameter
which enters in stellar computations. However, as already discussed
in Paper I, it is usually constrained by the requirement that
the solar radius 
and the observed colors of red giant stars have to be correctly
reproduced.
In Paper I we have demonstrated that the value of {\sl ml}
obtained by reproducing the $\rm T_{eff}$ of the Sun may not be suitable also for
reproducing the $\rm T_{eff}$ of the GCs RGB. For this reason we have
performed the {\sl ml} calibration for our evolutionary tracks of metal
poor low mass stars by reproducing the observational $\rm T_{eff}$
of GCs RGBs as derived by Frogel et al. (1983).

In order to illustrate the dependence of the bump
luminosity on the value of the mixing length, we have reexamined the
evolutionary tracks discussed in Paper I, computed for different
values of {\sl ml}. 
We find that ${\Delta\rm {V_{bump}}\over\Delta{ml}}\approx-0.27mag$.
Since we have verified that the level of the ZAHB in the RR Lyrae
stars region is not influenced
(as it is well known) by a variation of {\sl ml}, it is obtained that the value of
$\Delta\rm {V}^{bump}_{hb}$ is decreased by $\approx0.04$ mag for a variation by
+0.15 of {\sl ml}.

It is worth noting that a variation of {\sl ml} of about 0.15 is large
enough to compromise
the agreement between theoretical evolutionary models and the$\rm T_{eff}$ 
of GCs RGB stars. 
In Paper I we derived a relation between [M/H] and $\rm T_{eff}$ of
the GCs RGB by adopting the $\rm T_{eff}$ determinations  by Frogel et al. (1983);
this relation has been used for calibrating all the stellar models
presented in this paper. The dispersion of the observational
points around this relation is of the order of 100K (see the
discussion in Paper I), and since this dispersion corresponds to a 
variation of the {\sl ml} by around 
$\pm0.10$, we can assume this quantity as an estimate of the
maximum uncertainty associated to the calibrated value of {\sl ml}.
Taking into account this indetermination, 
the variation of the theoretical values of $\Delta\rm {V}^{bump}_{hb}$
results to be less than $\pm0.03$ mag, that is quite negligible.

\subsection{The age of the stellar population.}
\tx

The location of the bump on the RGB of a stellar track depends on
the mass of the model, i.e. on the age of the stellar system in which that
 star is now evolving (see Straniero \& Chieffi 1991), while the ZAHB
luminosity is practically independent on the age, at least in the age
range spanned by the GCs. 
From the evolutionary tracks of  
$0.75M_\odot$ - $0.8M_\odot$ - $0.9M_\odot$ models, we derive for each
metallicity an increase of
$\Delta\rm {V}^{bump}_{hb}$ by almost 0.024 mag for an increase of 1 Gyr in
the age of the clusters.  

\subsection{The effect of the mass loss on the RGB.}
\tx

As it is well known, all the low mass stars during their evolution along the RGB
experience the phenomenon the mass loss. Mass loss during the RGB
is also required to obtain the correct HB morphology
of GCs.
However mass loss does 
not affect at all the
main evolutionary properties of the stars, as for instance the helium
core mass at the he flash and the location in $\rm T_{eff}$ of the RGB.
This occurrence is really correct if "conventional" assumptions are made
on the efficiency of this phenomenon.
In fact, it has been shown (Castellani \& Castellani 1991,
D' Cruz et al. 1995) that, assuming a very high efficiency of the
mass loss mechanism, it could be possible to "obtain" stellar models whose 
evolution does not follow the prescriptions of the canonical stellar
 evolution theory.

We have tested if mass loss could affect in some way the bump
luminosity on the RGB.
For this aim, some evolutionary tracks have been computed adopting various
assumptions on the mass loss efficiency. Let us remember that the mass loss
phenomenon is usually parametrized in stellar computations 
the Reimers (1975) formula, in which appears a free parameter $\eta$.
According to various authors (see for instance Renzini \& Fusi Pecci 1988)
to finely reproduce the distribution of stars on the HB of the bulk of
galactic globular cluster - except the GCs showing HB blue tails in their 
CMDs it is necessary to adopt for such a parameter
a value around 0.3 - 0.4.
The evolutionary tracks corresponding to the same
$0.8M_\odot$ model computed without mass loss in a case and assuming a
large efficiency ($\eta=1.0$) for the mass loss are plotted in Fig.2.
\figure{2}{S}{100mm}{\bf Figure 2. \rm 
The H-R diagram for a 0.8$M_\odot$ stellar model computed under two
different assumptions concerning the efficiency of the mass loss
phenomenon. In the inset the effect of 
mass loss on the evolutionary tracks in the RGB bump region is shown.}
One can easily notice that the effect of the mass loss on the bump is 
absolutely negligible. It is also worth noticing  that this result has been
obtained assuming a very strong amount of mass loss, as it is confirmed by 
the occurrence that the star is forced to leave off the RGB before
igniting the He central burning.
Therefore one can safely assume that, in the range of efficiency of the mass
loss phenomenon in real RGB stars, the influence of this mechanism
on the bump luminosity is quite
negligible.

\subsection{The equation of state.}
\tx

A new equation of state suitable for stellar evolutionary
computations has been very recently provided by
Rogers, Swenson \& Iglesias (1996) (OPAL EOS).
Therefore we have decided to test if theoretical values of
$\Delta\rm {V}^{bump}_{hb}$ are modified when using this new physical
input in the evolutinary models computations. 
In this case we have supplemented
the OPAL EOS (in the regions not covered by the tables), with a Saha
EOS (for $T<5000K$) and the Straniero (1988) EOS, as described in
Salaris, Degl'Innocenti \& Weiss (1996).

The use of the OPAL EOS in evolutionary computations has led to a
revision of the age of the GCs with respect to determinations obtained
by means of stellar models computed using a different EOS (as in
Chaboyer et al. 1992 and Salaris et al. 1993).  
Recent works by Chaboyer \& Kim (1995, who find a reduction by 6-7$\%$
in the ages derived using $M_{V}(TO)$ and an average age  of 13 Gyr
for a sample of 40 clusters),
Mazzitelli, D'Antona \& Caloi
(1995), Salaris et al. (1996),
show that the average age
of the GCs should be around 12-13 Gyr, significantly lower than the average age obtained 
in previous works.
This result is very important since, as it is well known, the determination
of the GCs age is a fundamental tool to investigate the galactic 
formation mechanism and the age of the Universe. 
The discussion of these results is out of the goals of the present work, but
due to the effect of the age on the RGB bump location, we have
to make some realistic assumptions concerning the adopted value
for the age of the GCs.
Using the same criteria adopted to obtain our reference models,
in computing the theoretical values of $\Delta\rm {V}^{bump}_{hb}$
by using the OPAL EOS we have therefore chosen an average age of 12 Gyr for
the GCs.

Once calibrated the mixing length parameter on the observational data
by Frogel et al. (1983),  
we have computed a set of evolutionary tracks as described for our
reference case; ZAHB
luminosities result to be about 0.055 mag higher than for the models computed with the
Straniero EOS (see Fig.1) 
and $\Delta\rm {V}^{bump}_{hb}$ values are 
0.04-0.05 mag lower than the ones obtained adopting our reference scenario
and the Straniero EOS.
When adopting an age of 15 Gyr also for the
models computed with the OPAL EOS, 
$\Delta\rm {V}^{bump}_{hb}$
values 0.03 mag higher than the reference values are obtained.

\subsection{The opacity.}
\tx

The bump luminosity, being related to the position of the H 
discontinuity produced by the convective envelope during its deeper
penetration, depends strongly on the opacity evaluation for temperatures of
the order of one milion of degrees, i.e.
the temperature at the bottom of the convective envelope.
This occurrence has been tested in the past when changing the opacity tables
from the Cox \& Stewart (1970) and Cox \& Tabor (1976) to the most reliable
Los Alamos opacity library (hereinafter LAOL; Huebner et al. 1977) the brightness of the bump
decreased by $\simeq0.2$mag; the change from the LAOL to the OPAL
opacities causes a much smaller reduction of the bump luminosity, by   
$\simeq0.07$mag.
Now the opacity evaluations are more accurate than in the past, both
in the high temperature and in the low temperature region
(see e.g. Paper I for a comparison between three recent sets of low
temperature opacities). So one can be hopefully
confident that there is not much room for a significative variation in the bump
brightness as due to a forthcoming generation of updated opacity libraries.
For instance, in very recent time, a big effort has been made  to improve
the accuracy of the OPAL opacity evaluations, by increasing the number of 
elements taken into account in the metal mixture (Roger \& Iglesias 1995 -
21 elements mixture).
The effect of this last generation of the OPAL opacities on the bump luminosity
has been also checked. As a result, there are not significative variations
of $\Delta\rm {V}^{bump}_{hb}$.

\subsection{Bolometric corrections.}
\tx

Our theoretical evolutionary tracks have been transformed into the
observational plane by adopting the Kurucz (1992) transformations. In
order to check the sensitivity of $\Delta\rm{V}^{bump}_{hb}$ to different
sets of bolometric corrections adopted, we have also used the
transformations by Buser \& Kurucz (1992) supplemented with the Buser
\& Kurucz (1978) ones for $T>6000K$, as described in Salaris et al. (1996).
The same values of $\Delta\rm {V}^{bump}_{hb}$ -
within 0.01 mag - have been obtained at each metallicity.

\section{Theory {\sl versus} Observations.}

\tx

As discussed in the first section, RGB bumps have been detected only
in a few globular clusters. To perform a meaningful comparison between
theory and observations, we have to consider clusters with a quite
clear detection of the bump, and with an accurate determination of the
metallicity (taking into account also the enhancement of the $\alpha$-elements),
since the $\Delta\rm {V}^{bump}_{hb}$ is strongly dependent on
the adopted global metallicity. For this reason
only clusters with high resolution spectroscopical
determinations of the photospheric abundances of Fe and $\alpha$
elements have been taken into account.
In Tab.2 the seven clusters considered, the adopted values of [M/H] (these values come 
from Salaris \& Cassisi 1996, where determinations of [$\alpha$/Fe]
and [Fe/H] are collected for a sample of 22 globular clusters), the values
of $\rm V_{zahb}$, $\rm V_{bump}$, and $\Delta\rm {V}^{bump}_{hb}$ (with its associated
observational error) are reported. The sources of the 
photometric data are given in the discussion of the
individual clusters (see below). 
Five clusters of this GCs sample are very
poorly populated or not populated at all in the RR Lyrae stars region
(NGC6397, NGC6752, M79, 47 Tuc and NGC6352),
and the determination of the ZAHB luminosity in the
instability strip is
not straightforward.  
The values of $\rm V_{zahb}$ reported in Tab.2 have
been carefully determined by adopting different procedures, depending on the
morphology of the cluster HB.

In the case of M3 and M5, whose CMDs display a very well
populated HB in the RR Lyrae region, we have adopted  
the mean luminosity of the RR Lyrae variables ($\rm \langle V_{RR}\rangle $) 
as provided in the
original works (see below).
Clearly $\rm \langle V_{RR}\rangle $ 
does not represent the value of the ZAHB
luminosity, that has to correspond to the lower envelope
of the observed HB stars distribution (see e.g. Sandage 1990); 
thus, for obtaining  the ZAHB level one has to correct 
the $\rm \langle V_{RR}\rangle $ value taking into account the thickness of the observed HB.
Carney , Storm \& Jones (1992) provide a relation between $\rm V_{zahb}$
and $\rm \langle V_{RR}\rangle$ derived by using the data published by Sandage (1990),
who carefully studied the vertical distribution of HB stars in various GCs.
They used a sample of 8 clusters to derive the following relation:

$$\rm V_{zahb}=\langle V_{RR}\rangle +0.05[Fe/H]+0.20\,\,\,\,\,(4)$$

\noindent
that gives the ZAHB luminosity as a function of the mean luminosity
of the RR Lyrae stars and of the cluster iron content.
We have performed the same kind of analysis by using  clusters, 
in the sample considered by Carney et al. (1992),  for which spectroscopical 
determinations of [Fe/H] and $[\alpha/Fe]$ (see Paper I) are available,
thus obtaining the following relation:

$$\rm V_{ZAHB}=\langle V_{RR}\rangle+0.04[M/H]+0.15\,\,\,\,\,(5)$$

Relation (5) has been used in order to correct the values of
$\rm \langle V_{RR}\rangle $ for obtaining the ZAHB level. In each case
this formula has been used, we have verified that the
resulting luminosity matches
the lower envelope of the HB stellar population (see Fig. 3a,b). 

In the case of NGC6397, NGC6752 and M79 which are characterized by blue
HBs, we have followed the same procedure used by other authors (see
Buonanno et al. 1986, Alcaino et al. 1987, Ferraro et al. 1992) in
order to derive their ZAHB luminosity. We have considered a reference cluster
with the same metallicity but with an HB populated in the blue 
and in the RR Lyrae region, and its CMD has been shifted in such a
way that the blue part of its HB sequence overlaps the blue HB of the      
cluster we are studying, in order to form a unique sequence (see Fig. 3c-e).
In this way the ZAHB level of the cluster with the blue HB has been
obtained from the reference cluster, after the correction for the relative
luminosity shift.
When it has been selected as reference cluster a GC with the appropriate
metallicity, and if the CMD shift has been correctly done,
the two RGBs have also to be perfectly overlapped (see Fig. 3c-e), and the
horizontal shift applied to the reference cluster has to correspond to
the reddening difference between the two clusters.

As for 47 Tuc and NGC6352 
a procedure very similar to that described in Fullton et al. (1995)
has been adopted.
The HBs of these two clusters are populated only on the red side of the
instability strip, so no estimate of the ZAHB level at the
region of the RR-Lyrae stars is available.
Moreover, the value
of the ZAHB at the red side can not be used as an estimate
of the ZAHB luminosity for the RR Lyrae region
(see the discussion in Castellani, Chieffi \& Pulone 1991). 
In this case we have derived from our ZAHB models (transformed to
the observational plane by adopting both the Kurucz 1992 and the
Buser \& Kurucz 1978, 1992 transformations) for Z=0.003 and Z=0.006 
the difference $\delta$ in
$\rm M_{V}$ between the red part of the ZAHB and the point along the
ZAHB at
$\rm log(T_{eff})=3.85$.
We obtained $\delta=0.10\pm0.05$, and applied this
correction to the observational data, in order to perform the comparison
with the theoretical luminosities at $\rm log(T_{eff})=3.85$.

In the following the case of each cluster (in order of increasing
metallicity) will be discussed separately: 
\figure{3}{D}{180mm}{\bf Figure 3. \rm 
CMDs of the clusters for which 
the ZAHB luminosity level has been obtained using our relation (5), or
superimposing another cluster populated in the RR Lyrae region (see text).
In each panel
the adopted ZAHB luminosity level and the associated error is indicated.
In panel c) the C-M diagram of M68 (filled triangles) is superimposed to
that of NGC6397 (open circles); the same in panel d) and e), but with M3 (filled
triangles) superimposed respectively to NGC6752 and M79 (open circles).}
\table{2}{S}{\bf Table 2. \rm Luminosity of the bump,
of the Horizontal branch at $\rm log(T_{eff})=3.85$ and $\Delta\rm {V}^{bump}_{hb}$
for the clusters considered.} 
{\halign{%
\rm#\hfil&\hskip10pt\hfil\rm#\hfil&\hskip10pt\hfil\rm#\hfil&\hskip10pt\hfil\rm\hfil#\hfil&\hskip10pt\hfil\rm\hfil#\cr
Name & $\rm V_{zahb}$  & $\rm V_{bump}$ & $\Delta\rm {V}^{bump}_{hb}$ & [M/H] \cr 
\noalign{\vskip 10pt}
   NGC104 & 14.20 &  14.55 &  $\,\,\,\,0.35\pm0.18$ & -0.70  \cr
  NGC1904 & 16.36 &  16.00 &  $-0.36\pm0.12$ & -1.27  \cr
  NGC5272 & 15.76 &  15.40 &  $-0.36\pm0.07$ & -1.31  \cr
  NGC5904 & 15.15 &  14.95 &  $-0.20\pm0.07$ & -1.19  \cr
  NGC6352 & 15.50 &  15.86 &  $\,\,\,\,0.36\pm0.12$ & -0.70  \cr
  NGC6397 & 13.02 &  12.60 &  $-0.42\pm0.14$ & -1.70  \cr
  NGC6752 & 13.86 &  13.65 &  $-0.21\pm0.12$ & -1.28  \cr}}
\medskip
\noindent
{\bf (i) NGC6397}:  
The luminosity of the bump ($\rm V_{bump}=12.60\pm0.10$) 
is derived from FP90 who adopt the photometry 
by Alcaino et al. (1987). 
In the paper by Alcaino et al. (1987) the mean luminosity
of the HB in the RR-Lyrae stars region was obtained by superimposing
the CMD of M15 (which is populated in the RR Lyrae
region) to that of NGC6397 (see previous discussion). We have
followed the same procedure (see Fig. 3c) 
using the CMD of  M68 (Walker 1994), which has a global metal abundance
much more similar to that of NGC6397 
according to the latest spectroscopical determinations ([M/H]=-1.78,
see Paper I), and a well populated HB in the variable stars region. 
Walker (1994) provides $\rm \langle V_{RR}\rangle =15.64\pm0.01$ for the RR Lyrae
stars in M68, to which we have
subtracted a quantity $\Delta\rm {V}=-2.70$, corresponding to the shift we have
applied to the M68 diagram in order to match the blue HB of
NGC6397. The horizontal shift ($\Delta(B-V)$) that we applied to M68 is 
$\Delta\rm {(B-V)}$=+0.1, in good agreement with the reddening difference
between these two clusters ($\rm E(B-V)_{M68}=0.06-0.10$ according to
Walker 1994, and $\rm E(B-V)_{NGC6397}=0.17-0.20$ according to Alcaino
et al. 1987).  
Estimating an uncertainty of around 0.1 mag due to the procedure
previously described, after using relation (5) a value
$\rm V_{zahb}=13.02\pm0.10$ is finally derived.
\smallskip
\noindent
{\bf (ii) NGC5272 (M3)}:
The observational data come from Buonanno et al. (1994). In this paper
the authors provide the value of $\rm V_{bump}=15.40\pm0.05$ and
$\rm \langle V_{RR}\rangle =15.66\pm0.05$.
By applying the correction for the ZAHB we obtain $\rm V_{zahb}=15.76\pm0.05$.
\smallskip
\noindent
{\bf (iii) NGC6752}:
The bump level ($\rm V_{bump}=13.65\pm0.05$) comes from FP90 and has been 
determined by using the photometry by Buonanno et al. (1986). 
In order to derive the ZAHB level we have 
superimposed the data of M3 (which has a very well populated RR
Lyrae region) to that of NGC6752 (see Fig.3d ), since the
two clusters have quite the same metallicity (see Tab.2).
The ZAHB luminosity of M3 has been corrected by a quantity
$\Delta\rm {V}=-1.90$ corresponding to the vertical shift applied to
its CMD in order to match the data of NGC6752.
The horizontal shift $\Delta\rm {(B-V)}=+0.05$ applied to M3 agrees very
well with the reddening difference between the two clusters 
($\rm E(B-V)_{M3}=0.00-0.03$ according to
Buonanno et al.  1994, and $\rm E(B-V)_{NGC6752}=0.02-0.06$ according to Penny
\& Dickens 1986).  
We derive $\rm V_{zahb}=13.86\pm0.11$, after taking into account an
error of about 0.1 mag due to the procedure of superimposing the two
HB sequence, and the error associated to the estimate of the M3 ZAHB
level (see above). 

It is very interesting to note the agreement between the cluster distance
modulus obtained by adopting this value of $\rm V_{zahb}$ together
with  our
relation (2), and the distance modulus determined very recently by
Renzini et al. (1996) in a completely independent way.
They use the observed White Dwarf (WD) cooling
sequence of NGC6752 as a distance indicator, and derive the distance
modulus by fitting the cluster WD sequence to an empirical cooling
sequence constructed using local WDs with well determined
trigonometrical parallaxes. Following this procedure they derive 
$\rm (m-M)_{o}=13.05$ with an overall uncertainty less than $\pm0.1$ mag. 
By adopting our relation (2) together with$\rm V_{zahb}=13.86\pm0.11$,
and $\rm A_{V}=0.12\pm0.06$ (see Renzini et al. 1996, Penny \& Dickens
1986), we obtain  
$\rm (m-M)_{o}=13.01\pm0.13$, in good agreement with the result by
Renzini et al (1996).
\smallskip
\noindent
{\bf (iv) NGC1904 (M79)}:
The observational data come from Ferraro et al. (1992).
We adopt
$\rm V_{bump}=16.00\pm0.05$ as estimated by the authors.
From the observational data we have determined the ZAHB level following the same
procedure as described for NGC6397 and NGC6752. 
Also in this case we have shifted the data of M3 in order to superimpose the two 
HB sequences (see Fig.3e).
The vertical shift applied to M3 is $\Delta\rm {V}=+0.6$, while no
correction to its (B-V) values has been applied; this is in good
agreement with the fact that the reddening of the two clusters ($\rm
E(B-V)_{M79}=0.00-0.02$ according to Ferraro et al. 1992) is coincident. 
The ZAHB luminosity of M79 results to be $\rm V_{zahb}=16.36\pm0.11$.
\smallskip
\noindent
{\bf (v) NGC5904 (M5)}:
The bump luminosity has been derived from Brocato, Castellani 
\& Ripepi (1995, 1996) ($\rm V_{bump}=14.95\pm0.05$). 
From the same authors we have derived an average magnitude for the RR Lyrae stars 
$\rm \langle V_{RR}\rangle =15.05\pm0.05$.
By applying relation (5) $\rm V_{zahb}=15.15\pm0.05$ is obtained. 
\smallskip
\noindent
{\bf (vi) NGC104 (47 Tuc) and NGC6352}: 
The bump ($\rm V_{bump}=14.55\pm0.05$) and the ZAHB luminosities 
($\rm V_{zahb}=14.10\pm0.15$) of 47 Tuc come from FP90, who
adopted the photometry by King et al. (1985). As for NGC6352 we
have used the data from Sarajedini \& Norris (1994) obtaining
$\rm V_{zahb}=15.40\pm0.10$ from the fit of the lower envelope of the HB
stellar population,
and $\rm V_{bump}=15.86\pm0.05$ as quoted by the authors.
Fullton et al. (1995) published recently another photometry for
NGC6352, obtained in part from ground based observations and in part from $HST$
observations, and they found a zero point difference in the V magnitudes
between their work and the results by Sarajedini \& Norris, for the 145 stars
in common. However, this zero
point difference (around 0.16 mag) does not affect at all - as obvious
- the differential quantity $\Delta\rm {V}^{bump}_{hb}$. 

By applying the correction $\delta=0.10\pm0.05$ due to their red HB
(see the previous discussion), we obtain $\rm V_{zahb}=14.20\pm0.17$
for 47 Tuc
and $\rm V_{zahb}=15.50\pm0.11$ for NGC6352.

\smallskip

\figure{4}{S}{80mm}{\bf Figure 4. \rm 
The values of ${\Delta}\rm V^{bump}_{hb}$ {\it versus} the global metallicity
for all the clusters in our sample. Our {\it reference} relation (see text)
(solid line) and the theoretical prescription obtained
using in stellar computations the OPAL EOS (dashed line) are also plotted.}

In Fig.4, the comparison between the observational
values of $\Delta\rm {V}^{bump}_{hb}$ for
the seven clusters considered and the theoretical prescriptions is reported.
We have taken into account the observational errors in the determination of
such a quantity (see Tab. 2), and assumed an uncertainty of 0.15dex on the
spectroscopic estimates of [M/H] (see the discussions in Gratton,
Quarta \& Ortolani 1986, Gratton \& Ortolani 1989 and Kraft, Sneden,
Langer \& Shetrone 1993).
The theoretical relations corresponding to the models computed 
adopting the Straniero (1988) EOS and the OPAL EOS are
displayed. These two relations correspond - as previously discussed -
to an age of 15 Gyr (Straniero EOS) and 12 Gyr (OPAL EOS),
and they have been computed by adopting Y=0.23.

The figure clearly demonstrates the overall agreement between
theoretical standard stellar models (computed with the Straniero or the OPAL EOS)
and observations, either for the
run of  $\Delta\rm {V}^{bump}_{hb}$ with respect to [M/H], either for the
absolute values of this quantity; all the
observational points are fitted within the observational error bars. 
The worst agreement is obtained for NGC6397, the most metal poor
cluster in our sample. 

We may therefore conclude that  
{\sl there is no significant discrepancy
between observations and canonical stellar models} computed by
adopting updated input physics, under the assumption of a constant age
for the GCs, and a constant initial Helium abundance.

\section{DISCUSSION AND CONCLUSIONS.}

\tx

In the two previous sections we have presented new theoretical stellar models
computed with updated input physics, with the aim of comparing the
theoretical values of $\Delta\rm {V}^{bump}_{hb}$ with the observations.
Before comparing this quantity with  real clusters, we had to make
some assumptions about the age and the original He content of our models,
since the
$\Delta\rm {V}^{bump}_{hb}$ is influenced both by the age (a change
of the age affects the bump luminosity) and by the value of Y (a
variation of the Helium content modifies the ZAHB and bump luminosities). 
In the conservative hypothesis of coeval GCs with the same
original He content (Y=0.23), a very good agreement
between theory and observations can be derived from Fig.4.
This means that standard stellar
models can actually reproduce the luminosity levels of ZAHB and bump,
and their run with respect to the metallicity; it confirms also that 
standard stellar models reproduce accurately the Hydrogen profile in the
interior of RGB GCs stars.
\figure{5}{S}{80mm}{\bf Figure 5. \rm 
As in Fig.4, but the theoretical relation $\Delta\rm {V}^{bump}_{hb}$ -
[M/H] are now displayed for our {\it reference} case (solid line),
with two different assumptions concerning the cluster age (13Gyr - short
dashed line; 17 Gyr - long dashed line) and for the case of an
Helium abundance scaling with the metallicity (see text) (dotted line).} 
The differences between our results and the conclusions by FP90 are due
basically to three different reasons:
\smallskip\noindent
i) the use of updated evolutionary models;
\smallskip\noindent
ii) the adoption of new spectroscopical determinations of $[\alpha/Fe]$
and [Fe/H]; 
\smallskip\noindent
iii) new observational data.
\smallskip\noindent
FP90 have used models by Rood computed with the old Cox \& Stewart
(1970) and Cox \& Tabor (1976) opacities and, as discussed in their 
paper, it is not taken into account in the computation of their HB models 
the extra helium brought to the surface during the first dredge up.    
In addition, as yet discussed in the introduction, they used also a
different 
set of bolometric corrections
(Bell \& Gustafsson 1978 and Kurucz 1979).
The net resulting difference between our theoretical values of $\Delta\rm
{V}^{bump}_{hb}$ and the ones used by FP90 (that arises from the
differences on both the ZAHB and the Bump levels)
ranges from 0.12 up to 0.22
mag (it is higher at higher metallicities). 
The remaining part of the discrepancy between theory and observations
disappears when considering individual high resolution 
spectroscopical determination of [M/H],
including the overabundance of the $\alpha$ elements, and adopting
(when they are available) more recent observational data.

As a further step
we can now check how the fit presented in Fig.4 is modified by
relaxing the hypothesis of coeval GCs or of a constant Y. If we
consider a constant Y but an age spread by $\pm2$ Gyr around our
assumed average age of the clusters (see e.g. Chaboyer, Demarque \&
Sarajedini 1996), the theoretical $\Delta\rm {V}^{bump}_{hb}$ 
values would be spread by approximately $\pm0.05$ mag around the
line displayed in Fig.4. In Fig.5 two lines corresponding to the
$\Delta\rm {V}^{bump}_{hb}$ values obtained for 
our average age increased and decreased by 2 Gyr are displayed.
As for a variation of Y with the metallicity, 
the effect of a $dY/dZ=3$ on the $\Delta\rm {V}^{bump}_{hb}$ is shown
in Fig.5.
It is evident from the figure that the adopted variation of Y with the
metallicity has a negligible effect on the theoretical
$\Delta\rm {V}^{bump}_{hb}$ values 
(at most 0.02 mag at [M/H]=-0.6), and also the assumed spread in the ages of the
clusters does not change very much the theoretical $\Delta\rm {V}^{bump}_{hb}$-[M/H]
relation, remaining always compatible with the observational data.
Because the agreement between our theoretical $\Delta\rm {V}^{bump}_{hb}$
values  and the observations,
relation (3)  can be safely used for deriving the clusters metallicities. 
By simply differentiating this relation, one obtains that a
variation of the metallicity by $\pm0.15$ dex around [M/H]=$-1.3$ 
corresponds to a variation of
$\Delta\rm {V}^{bump}_{hb}$ by $\pm0.12$ mag. 
Allowing for an observational error of around 0.12 mag (see for
example Tab.2) in the
determination of $\Delta\rm {V}^{bump}_{hb}$, and for an error of around 0.1
mag due to uncertainties in the age, He content of the cluster and the
mixing length calibration of the theoretical models (see section 2), it
would be possible to estimate the metallicity of the cluster with an
error of around $\pm0.20$ dex (if [M/H] is around -1.3).
It is worth noticing that the luminosity difference between the bump and the
Horizontal branch is a metallicity index which relies only on the observation 
of luminous stars in the GCs and  is 
less dependent on the precise value of the mixing length than others
metallicity indicators based on the colors or on the shape of the RGB.

When the metallicity of a cluster is known, it is possible to use the
bump luminosity as a standard candle. Relation 
(3) can be used as a tentative guess
to estimate the HB luminosity level in those globular clusters 
in which the horizontal branch is poorly populated near the RR-Lyrae
instability strip or the HB is quite blue or quite red.
It could be interesting to test this procedure in evaluating the Helium
abundance on the basis of the R method in those clusters for which estimating the HB
luminosity level is a thorny problem. 

Before concluding we want to stress again that in this investigation
we have computed canonical stellar evolutionary models, neglecting
other non canonical effects, as, for instance, the Helium (and heavy
elements) diffusion. The inclusion of this mechanism in solar models   
changes, among other quantities, the chemical abundances in the
convective envelope and the depth of the
convective region with respect to canonical models, improving the
agreement with helioseismological data (see, e.g., the discussion in
Castellani et al. 1996). 

Proffit \& Vandenberg (1991) have studied the evolution of GCs stars
taking into account He diffusion, and very recently Castellani et al
(1996) have shown that He and heavy elements diffusion does not change
appreciably (age differences by less than 1 Gyr)
the age of the GCs with respect to canonical evaluations,
but at present time
an analysis of the influence of the He and heavy elements
diffusion on the $\Delta\rm {V}^{bump}_{hb}$ does not yet exist;
from preliminary computations at Z=0.0004 (Degl'Innocenti 1996,
private communication) it results that this quantity is changed by
only $\approx 0.02$ mag with respect to the canonical value, but
in order to test further the stellar models which include diffusion,
computations also for
higher metallicities have to be performed.

\section*{Acknowledgments}

\tx 
We gratefully thank Giuseppe Bono, Vittorio Castellani,
Martijn De Kool and Achim Weiss
for helpful discussions and a careful preliminary reading of the manuscript;
Scilla Degl'Innocenti, Anna Piersimoni and Oscar Straniero
are acknowledged for stimulating discussions on this subject.
We also warmly thank Dave Alexander for providing us with his $\alpha$-enhanced
low temperature molecular opacities, and the the referee, Brian Chaboyer,
for constructive remarks and observations
that have improved the level of the paper.

\section*{References}

\bibitem Alcaino G., Buonanno R., Caloi V., Castellani V., Corsi C.E.,
Iannicola G. \& Liller W. 1987, AJ 94, 917
\bibitem Alexander D.R. \& Ferguson J.W. 1994, ApJ 437, 879
\bibitem Alongi M., Bertelli G., Bressan A. \& Chiosi C. 1991, A\&A 244, 95
\bibitem Bahcall J.N. \& Loeb A. 1990, ApJ 360, 267
\bibitem Bell R.A. \& Gustafsson B. 1978, A\&AS 34, 229
\bibitem Bergbush P.A. 1993, AJ 106, 1024
\bibitem Bono G. \& Castellani V. 1992, A\&A 258, 385
\bibitem Brocato E., Buonanno R., Malakhova Y. \& Piersimoni A.M. 1996, A\&A {\it in press}
\bibitem Brocato E., Castellani V. \& Ripepi V. 1995, AJ 109, 1670
\bibitem Brocato E., Castellani V. \& Ripepi V. 1996, AJ 111, 809
\bibitem Buonanno R., Caloi V., Castellani V., Corsi C.E., Fusi Pecci
F. \& Gratton R.G. 1986 A\&AS 66, 79
\bibitem Buonanno R., Corsi C.E., Buzzoni A., Cacciari C., Ferraro F.R. \&
Fusi Pecci F. 1994, A\&A 290, 69
\bibitem Buser R. \& Kurucz R.L. 1978, A\&A 70, 555
\bibitem Buser R. \& Kurucz R.L. 1992, A\&A 264, 557
\bibitem Caloi V., Castellani V. \& Tornamb\'e A. 1978, A\&AS 33, 169
\bibitem Carney B.W., Storm J. \& Jones R.V. 1992, ApJ 386, 663
\bibitem Castellani M. \& Castellani V. 1993, ApJ 407, 649
\bibitem Castellani V., Chieffi A. \& Norci L. 1989, A\&A 216, 62
\bibitem Castellani V., Chieffi A. \& Pulone L. 1991, ApJS 76, 911
\bibitem Castellani, V., Ciacio F., Degl'Innocenti S. \& Fiorentini
G. 1996, preprint
\bibitem Chaboyer B. \& Kim Y.-C. 1995, ApJ 454, 767
\bibitem Chaboyer B., Demarque P. \& Sarajedini A. 1996, ApJ 459, 558
\bibitem Chaboyer D., Sarajedini A. \& Demarque P. 1992, ApJ 394, 515
\bibitem Chaboyer B., Kernan P.J., Krauss L.M. \& Demarque P. 1995,
preprint CITA-95-18
\bibitem Chieffi A. \& Straniero O. 1989, ApJS 71, 47
\bibitem Chieffi A., Straniero 0. \& Salaris M. 1995, ApJL 445, 39
\bibitem Clementini G., Carretta E., Gratton R., Merighi R.,
Mould J.R. \& McCarthy J.K. 1995, AJ 110, 2319
\bibitem Cox A.N. \& Stewart J.N. 1970, ApJS 19, 243
\bibitem Cox A.N. \& Tabor J.E. 1976, ApJS 31, 271
\bibitem D'Cruz N., Dorman B., Rood R. \& O'Connell R. 1995, Bull. American
Astron. Soc. 186, 2201
\bibitem Dorman B., Rood R.T. \& O'Connell R.W. 1993, ApJ 419, 596
\bibitem Ferraro F.R. 1992, MemSAIt 63, 491
\bibitem Ferraro F.R., Clementini, G., Fusi Pecci F. \& Sortino
R. 1992, MNRAS 256, 391
\bibitem Frogel J.A., Persson S.E. \& Cohen J.G. 1983, ApJS 53, 713
\bibitem Fullton L.K., Carney B.W., Olzewski E.W., Zinn R., Demarque
P., Janes K.A., Da Costa G.S. \& Seitzer P. 1995, AJ 110, 652
\bibitem Fusi Pecci F., Ferraro F.R., Crocker D.A., Rood R.T. \& 
Buonanno R. 1990,A\&A 238,95
\bibitem Gratton R.L. \& Ortolani S. 1989, A\&A 211, 41
\bibitem Gratton R.L., Quarta M.L. \& Ortolani S. 1986, A\&A 169, 208
\bibitem Grevesse N. 1991, in ``Evolution of stars: the photospheric 
abundance connection'', IAU Symp. eds. Michaud G., Tutukov A., p.63
\bibitem Huebner W.F., Merts A.L., Magee Jr. N.H.\& Argo M.F. 1977,
Los Alamos Sci. Lab. Rep. LA-6760-M
\bibitem Iben I. Jr 1968, Nature 220, 143
\bibitem Iglesias C.A., Rogers F.J. \& Wilson B.G. 1992, ApJ 397, 717
\bibitem Itoh N., Mitake S., Iyetomi H. \& Ichimaru S. 1983, ApJ 273, 774
\bibitem King C.R., Da Costa G.S. \& Demarque P. 1985, ApJ 299, 674
\bibitem Kraft R.P., Sneden C., Langer G.E. \& Shethrone M.D. 1993, AJ
106, 1490
\bibitem Kurucz R.L. 1979, ApJS 40, 1
\bibitem Kurucz R.L. 1992, in Barbuy B., Renzini A. (eds.), IAU
Symp. n. 149, ``The Stellar Populations of Galaxies'', Kluwer,
Dordrecht, p. 225
\bibitem Mazzitelli I., D'Antona F. \& Caloi V. 1995, A\&A 302, 382
\bibitem Peimbert M. \& Torres-Peimbert S. 1977, MNRAS 179, 217
\bibitem Penny A.J. \& Dickens R.J. 1986, MNRAS 220, 845
\bibitem Proffitt C.R. \& Vandenberg D.A. 1991, ApJS 77, 473
\bibitem Reimers D. 1975, Mem.Soc.Roy.Sci.Liege, $6^{e}$ Ser. 8, 369
\bibitem Renzini A. \& Fusi Pecci F. 1988, ARA\&A 26, 199
\bibitem Renzini A., Bragaglia A., Ferraro F.R., Gilmozzi R., Ortolani
S., Holberg J.B., Liebert J., Wesemael F. \& Bohlin R.C. 1996, ApJ in press
\bibitem Rogers F.J. \& Iglesias C.A. 1992, ApJS 79, 507
\bibitem Rogers F.J. \& Iglesias C.A. 1995, in Adelman S.J., Wiesse
W.L. (eds.), ``Astrophysical application of powerful new databases''
ASP Conference series, vol.78, p.31 
\bibitem Rogers F.J., Swenson F.J. \& Iglesias C.A. 1996, ApJ 456, 902
\bibitem Rood R.T. \& Crocker D.A. 1989, in Schmidt E.G. (ed.), IAU
Coll. 111, ``The use of Pulsating stars in Fundamental Problems of 
Astronomy'', Cambridge University Press, p. 103
\bibitem Salaris, Chieffi A. \& Straniero O. 1993, ApJ 414, 580
\bibitem Salaris M. \& Cassisi S. 1996, A\&A 305, 858
\bibitem Salaris M., Degl'Innocenti S. \& Weiss A. 1996, ApJ submitted 
\bibitem Sandage A. 1990, ApJ 350, 603
\bibitem Sarajedini A. \& Norris J.E. 1994, ApJS 93, 161
\bibitem Sarajedini A. \& Forrester W.L. 1995, AJ 109, 1112
\bibitem Straniero O. 1988, A\&AS 76, 157
\bibitem Straniero O. \& Chieffi A. 1991, ApJS 76, 525
\bibitem Straniero O., Chieffi A. \& Salaris M. 1992, MemSAIt 63, 315
\bibitem Thomas H.-C. 1967, Z.Ap. 67, 420
\bibitem Walker A.R. 1994, AJ 108, 555
\bibitem Wheeler J.C., Sneden C. \& Truran J.W. 1989, ARA\&A 27, 279

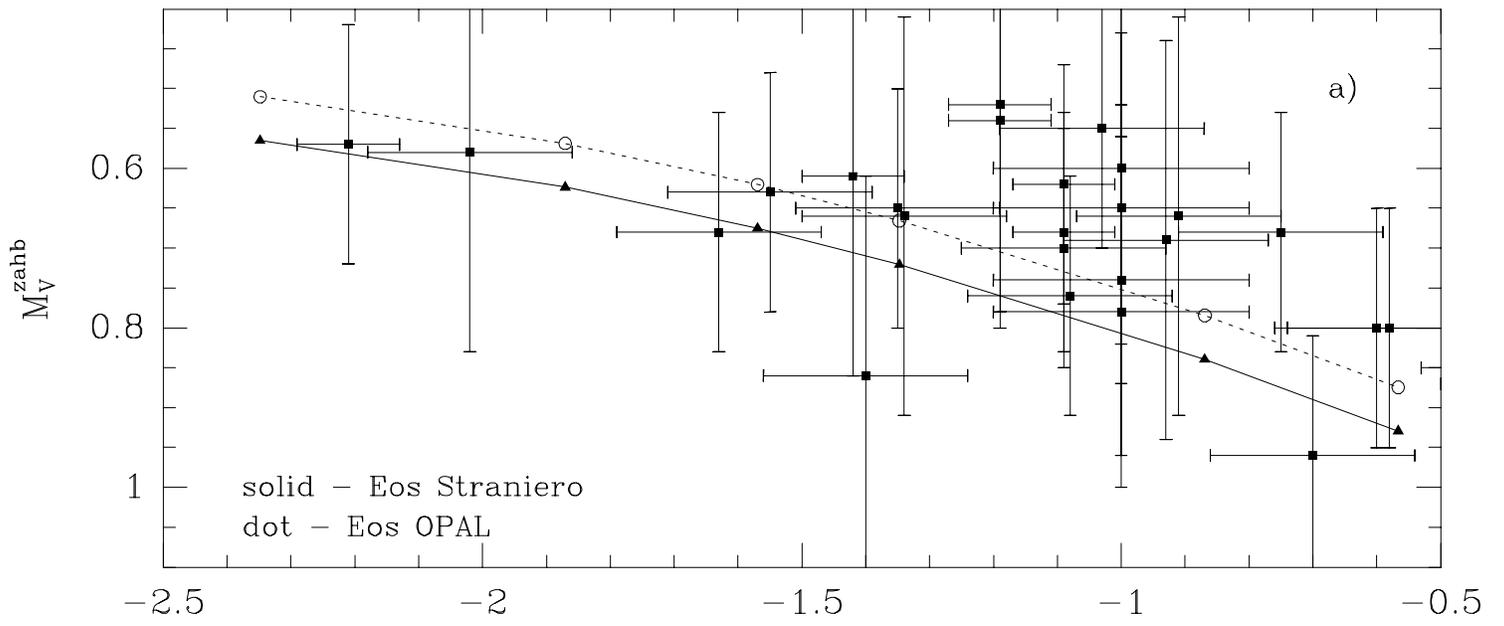
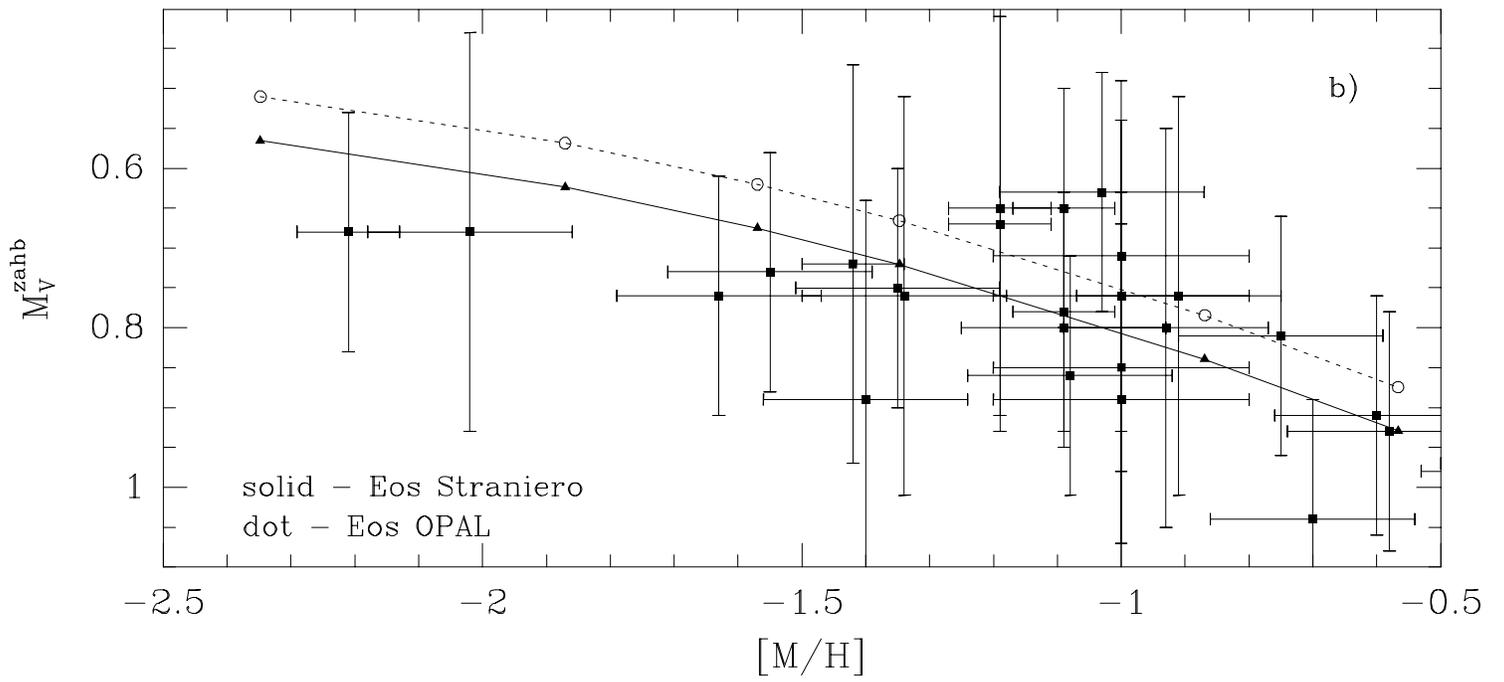
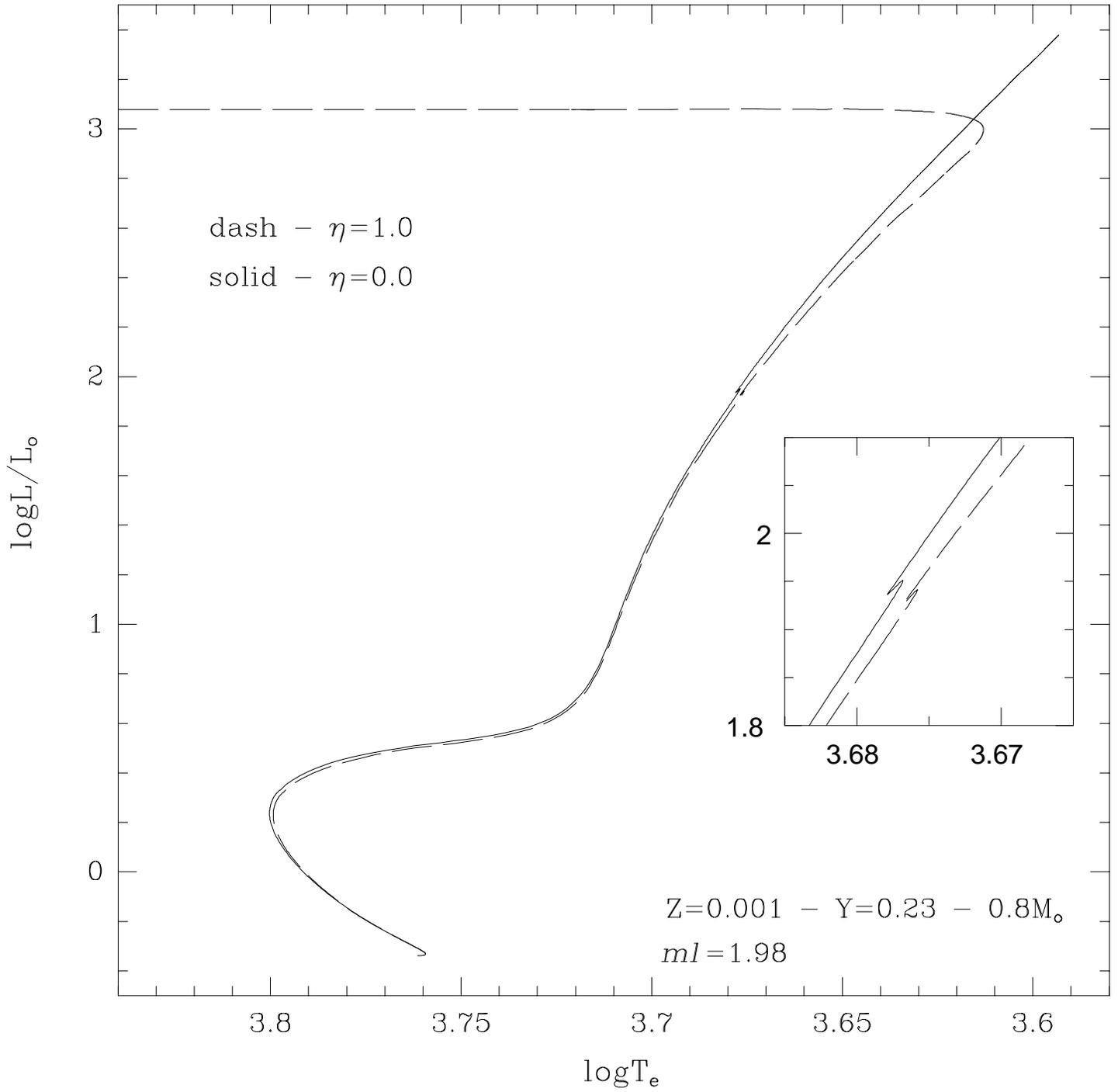
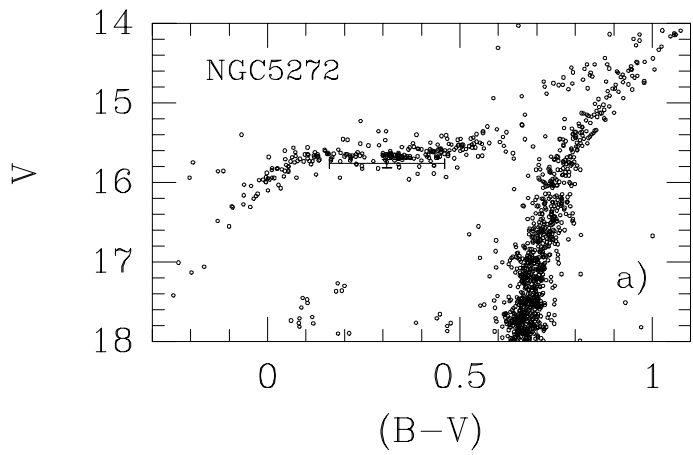
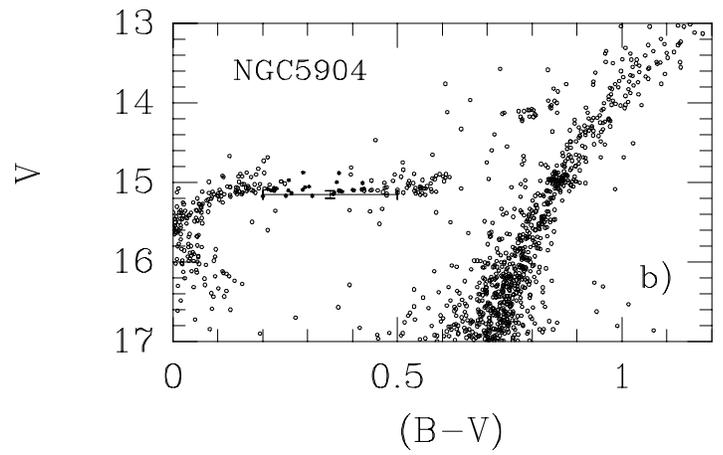
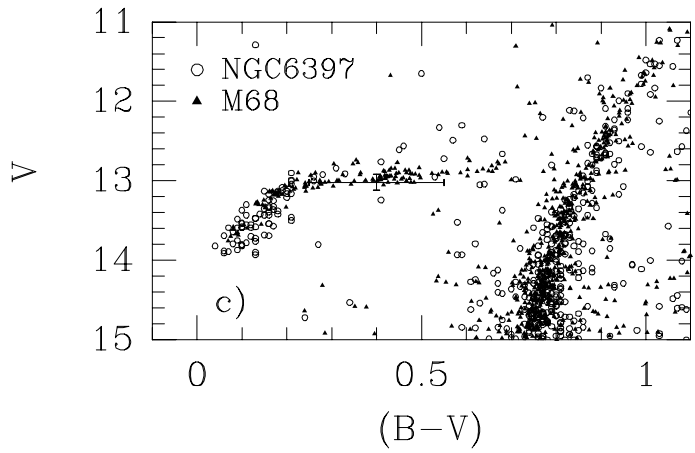
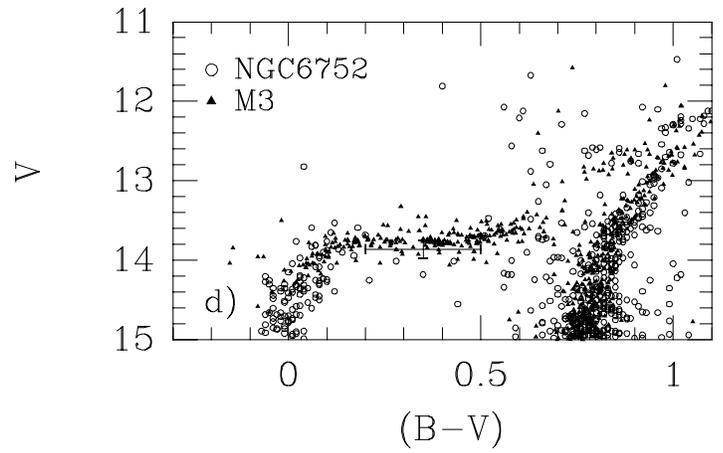
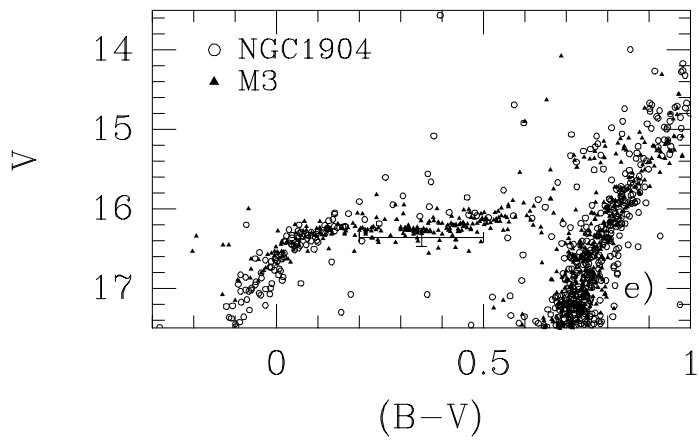
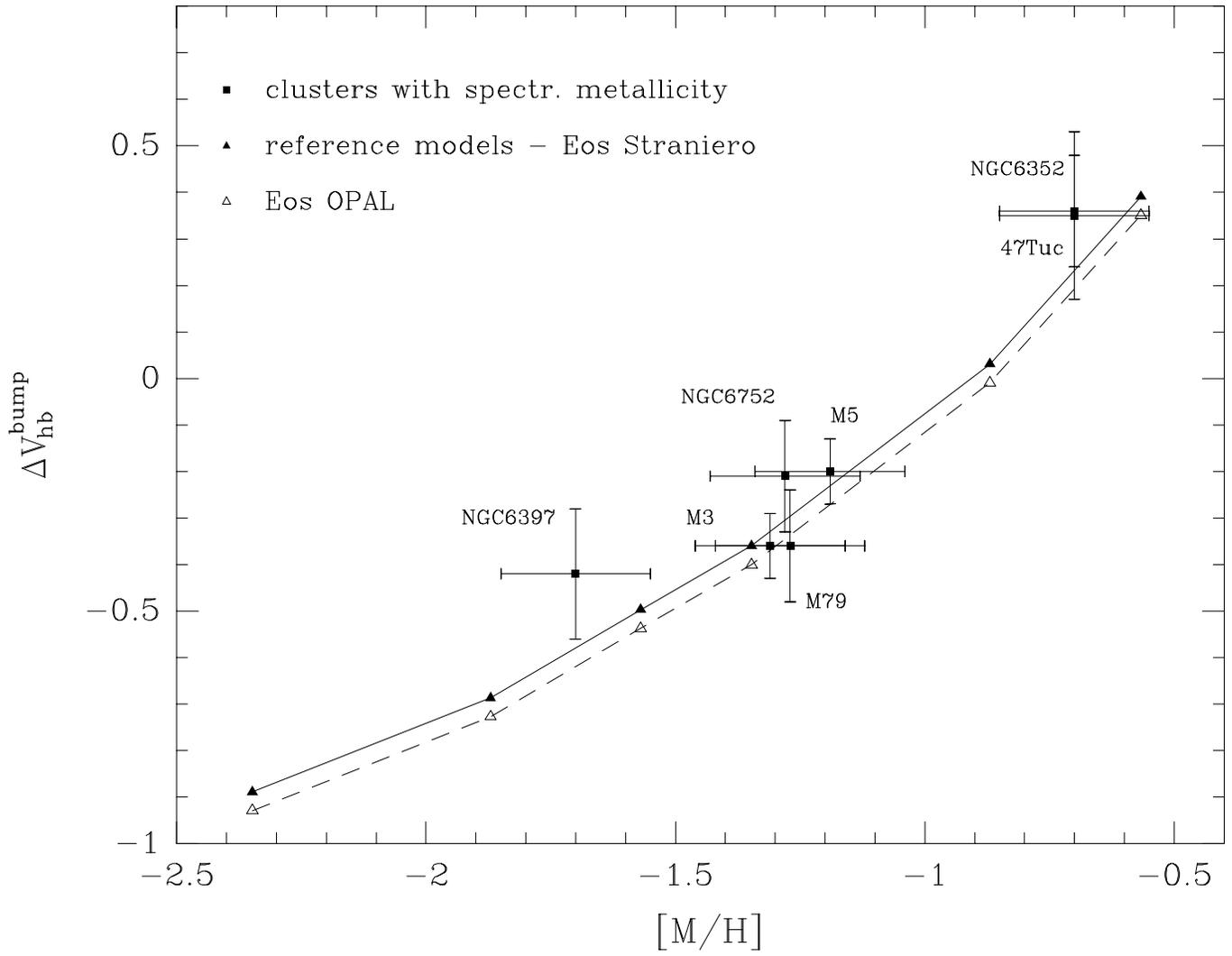
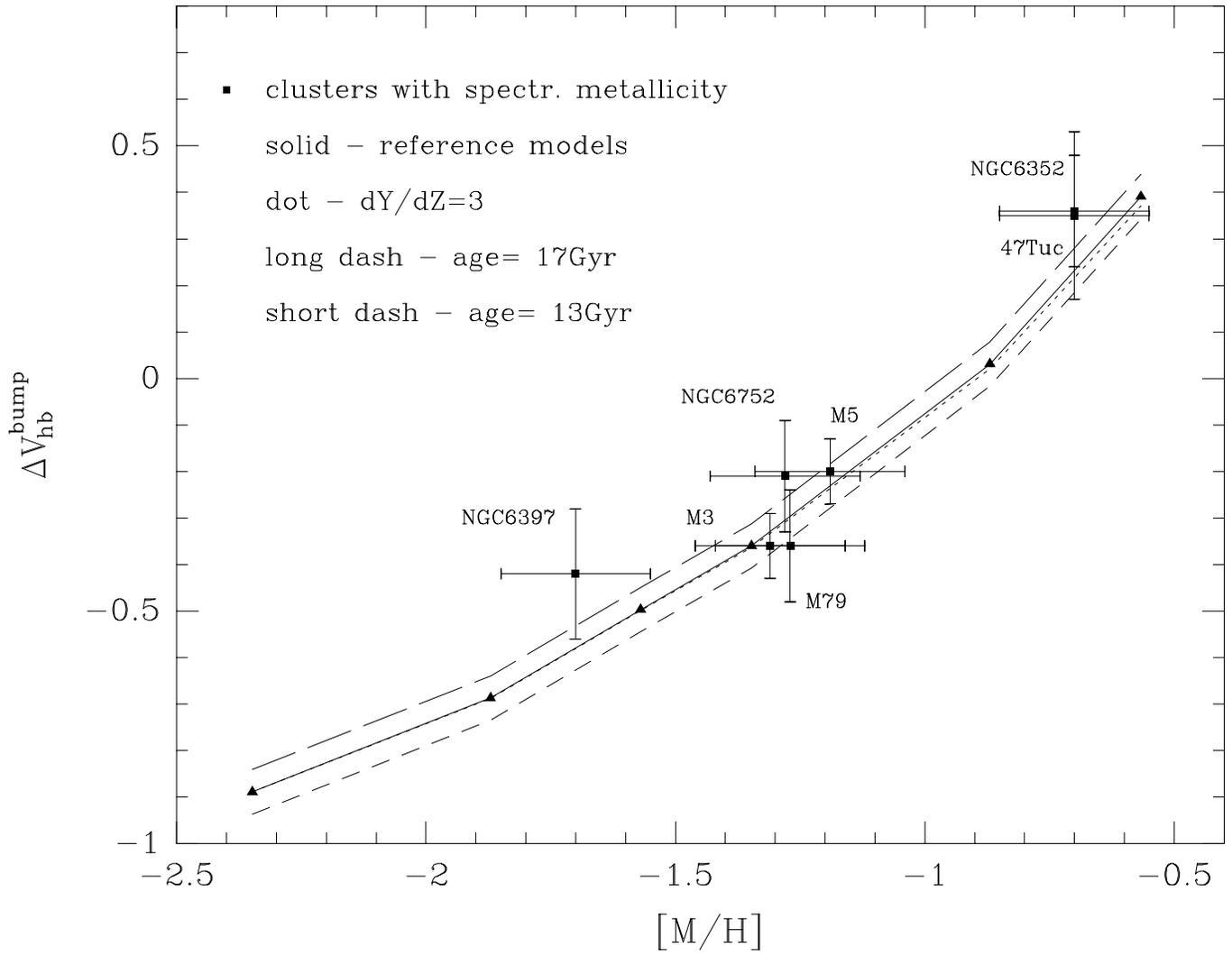
\bye